\documentclass[aps,twocolumn,amsmath,amssymb,floatfix,showkeys,
groupedaddress]{revtex4}

\usepackage{bm,amsmath,amssymb,amsfonts}

\usepackage[dvips]{graphicx}
\usepackage[colorlinks=true,citecolor=magenta,urlcolor=blue,filecolor=blue]{hyperref}
\usepackage{array}

\usepackage{color}
\definecolor{brown}{rgb}{0.70,0.30,0.05}
\begin{document}

\title{\textcolor{red}
{Engineering electronic states of periodic and quasiperiodic chains by buckling}}

\author{Amrita Mukherjee}
\email{amritaphy92@gmail.com}

\author{Atanu Nandy}
\email{atanunandy1989@gmail.com}

\author{Arunava Chakrabarti}
\email{arunava_chakrabarti@yahoo.co.in}

\affiliation{Department of Physics, University of Kalyani, Kalyani, 
West Bengal-741235, India}

\begin{abstract}
The spectrum of spinless, non-interacting electrons on a
linear chain that is buckled in a non-uniform, quasiperiodic manner  
is investigated within a tight binding formalism.
We have addressed two specific cases, viz., a perfectly periodic chain wrinkled 
in a quasiperiodic Fibonacci pattern, and a quasiperiodic Fibonacci chain, where the 
buckling also takes place in a Fibonacci pattern. The buckling brings distant neighbors 
in the parent chain to close proximity, which is simulated by a tunnel hopping 
amplitude. It is seen that, in the perfectly ordered case, increasing the strength 
of the tunnel hopping (that is, bending the segments more)  
absolutely  
continuous density of states is retained towards the edges 
of the band, while the central portion becomes fragmented 
and host subbands of narrowing widths containing extended, 
current carrying states, and multiple 
isolated bound states formed as a result of the bending. A 
switching ``on" and ``off" of the electronic transmission can thus
be engineered by buckling.
On the other hand, in the second example of a quasiperiodic Fibonacci chain, 
imparting a quasiperiodic buckling 
is found to generate {\it continuous subband(s)} destroying the usual 
multifractality of the energy 
spectrum. We present exact results based on a real space renormalization group 
analysis, that is corroborated by explicit calculation of the two terminal 
electronic transport.
\end{abstract}


\maketitle
\section{Introduction}
\label{intro}
Localization of single particle quantum states has been an ubiquitous phenomenon, 
observed primarily as a consequence of disorder in a system, and is traditionally 
called the Anderson localization~\cite{anderson}. The pivotal result in the field 
of disorder-induced localization is that, the single particle eigenstates of a 
Hamiltonian describing a disordered lattice should be exponentially localized 
for dimension $d \le 2$, and even for $d > 2$ for strong disorder. 
The effect is strongest in one dimension where all the states are localized 
for any strength of disorder. The envelope of 
the wave function decays exponentially with respect to a given location in the lattice
~\cite{kramer,abrahams}.
These results have been aptly justified by various calculations 
related to the localization length~\cite{rudo1,rudo2} and the density of 
states~\cite{alberto}. 
The single parameter scaling 
hypothesis -- its validity~\cite{rudo3}, variance~\cite{deych}, or even 
violation~\cite{bunde,titov} in low dimensional systems within a tight-binding 
approximation has also enriched the field.

This picture is to be contrasted with the effect of quasiperiodic order
~\cite{kohmoto1,kohmoto2} in one dimensional lattices where the single 
particle excitations are {\it critical} and exhibit a power law localization 
with a multifractal character in general ~\cite{mace}. Resistance for such systems exhibits 
power law growth as well as the size of the lattice 
increases~\cite{kohmoto3,sankar, macia}. The exotic spectral properties of 
quasiperiodic lattices occupied an immense volume of literature over the 
past three decades, and even today, the unusual behaviour of quantum conductance 
in such systems are of immense inetrest~\cite{varma}.

In this communication, we revisit the effect of 
`deterministic disorder', introduced in an
infinitely long one dimensional chain of atomic sites by buckling the chain 
in {\it local segments}, and throughout its 
length, following a quasiperiodically ordered sequence. 
Bent quantum wires have been studied previously in the context of 
ballistic transport characteristics~\cite{vacek}. Single and multiply bent 
two dimensional quantum wires were examined in respect of localized, doubly 
split one electron states~\cite{vakhnenko}. Apart from these, 
winding chains have been considered as models of 
conducting polymers~\cite{xiong1,xiong2}. The winding brings sites that were distant 
neighbors in the unperturbed system to close proximity, and a tunnel hopping 
provides additional paths for the electron. 
The lattice becomes a topologically   
disordered (deterministic though) system in the spirit of Guinea and 
Verg\'{e}s~\cite{guinea}, who discussed the effect of fluctuating coordination 
numbers in a linear chain, caused by dangling branches coupled from a side, or 
bridging two distant sites in a one dimensional lattice.
Such tunneling has been shown~\cite{xiong1,xiong2} 
to have profound influence on quantum transport, leading to large localization 
lengths in disordered polymers, which 
is indicative of a metal-insulator (MI) transition in such systems. 
Real polymers of course, include interaction between different constituent 
chains, the coupling between the adjacent chains being mediated by the 
overlapping hydrogen orbitals. In several cases the main spectral features 
can be explained by an effective tight binding model with next nearest neighbor 
hopping integrals. 
The effect of second neighbor hopping is also studied for quasi one dimensional 
organic polymer ferromagnetic systems and have unravelled a plethora of information
~\cite{kailun}. 
Buckling a chain makes such longer range interactions possible in a natural way, 
and thus deserves close scrutiny.

In addition, a quasiperiodic order in chains and their bending are two important 
ingredients that have been shown to arise in the recent field of two dimensional materials, 
a graphene nanoribbon for example~\cite{naumis}. This provides an extra motivation for studying the present model.

We examine two different cases. At first, we consider a perfectly periodic 
chain of atoms within a tight binding approximation. 
Buckling is introduced in a Fibonacci quasiperiodic sequence~\cite{kohmoto1} 
such that the chain bends after every $n$ and $m$ atoms, the numbers $n$ and 
$m$ being distributed in a Fibonacci pattern. Tunnel hopping is introduced across 
such clusters and it is seen that for large strength of the tunnel hopping amplitude, 
which may be thought to be caused by large bending of the local segments, the 
energy spectrum turns out to be extremely interesting. The outer edges of the spectrum 
retain the character of a pure, one dimensional chain of atoms, while the 
central part, spanning between $E=\epsilon \pm 2t$, feels the `disorder' and breaks up 
into multiple subbands populated by extended eigenstates as well as sharply 
localized bound states. Thus wrinkling the chain appropriately one can look into the 
possibility of a switching action as one sweeps the 
Fermi energy from the domain of absolutely continuous spectrum to the  
fragmented one, mixed with transparent and localized states.
\begin{figure}[ht]
\centering
\includegraphics[clip,width=6.5cm,angle=0]{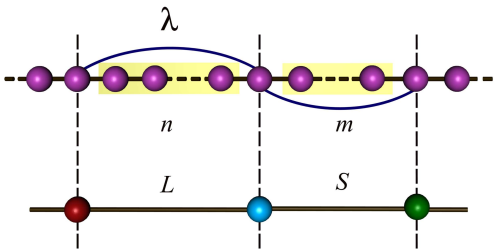}
\caption{(Color online) A one-dimensional chain of atoms with identical
quantum dots each of having on-site 
potentials $\epsilon$ and hopping integral $t$ with long range hopping 
$\lambda$ (blue line) across $n$-atoms and across 
$m$-atoms which follow a Fibonacci 
sequence, as discussed in text.
The figure below indicates the \textit{renormalized} version of
the buckled ordered chain with restricted long range hopping.} 
\label{order}
\end{figure}

In the second example, we consider a quasiperiodic Fibonacci chain to begin with. 
The tunnel hopping, spanning the second neighbors, but in a restricted sense, 
also follows a Fibonacci sequence. This introduces a competing quasiperiodic order. 
The introduction of the tunnel hopping in this case results in something 
totally different. We have encountered several cases where an appropriate choice of the 
tunnel hopping is seen to generate {\it absolutely continuous} subbands in the 
otherwise fragmented Cantor set energy spectrum, typical of a Fibonacci 
lattice~\cite{kohmoto1} making the system conduct over specified energy intervals.

In what follows, we discuss our findings in details. In section II we lay down the 
models and the methods followed. Section III contains the results for both the cases 
addressed here, and in section IV we draw our conclusions.

\section{The model and the method}
\subsection{The perfectly periodic chain with quasiperiodic buckling}
Let us refer to Fig.~\ref{order} which shows a linear chain of atomic 
scatterers (violet spheres). The chain is assumed to be buckled after
every $n$ and $m$ atoms, where we have taken the sequence of $n$ and $m$ 
following a Fibonacci distribution. That is, the chain is inhomogeneously distorted 
and the distorted segments are distributed as, $na$, $ma$, $na$, $na$, $ma$, $.......$, 
where, $a$ is the uniform lattice spacing. 
This is the typical arrangement of constituents in a binary 
Fibonacci chain comprising of say, two letters $L$ ans $S$, and grown following the 
algorithm $L \rightarrow LS$ and $S \rightarrow L$~\cite{kohmoto1}.
The Hamiltonian describing the system 
and written in a tight binding approximation reads,
\begin{equation}
H = \epsilon \sum_{i} |i\rangle \langle i| + \sum_{ij} t_{ij} |i\rangle \langle j|
\label{ham}
\end{equation}
In this case, 
$\epsilon$ is the uniform on-site potential describing 
the parent ordered chain. The hopping integral $t_{ij}=t$ for the 
nearest neighboring sites on the linear backbone, and $t_{ij}=\lambda_n$ or 
$\lambda_m$ depending on whether the buckling (shown by the blue line), 
connecting the vertices along the chain and  
across a segment of $n$ sites or a segment of $m$ sites. However, in the 
subsequent discussion, we shall stick to the case where, $\lambda_n = \lambda_m = 
\lambda$, which shows interesting spectral behavior.

Using a set of difference equations, 
\begin{equation}
(E - \epsilon) \psi_i = \sum_{ij} t_{ij} \psi_j 
\end{equation}
It is simple to reduce the original chain depicted in Fig.~\ref{order} 
to a linear Fibonacci chain with 
two kinds of (effective) bonds $L$ and $S$ (see Fig.~\ref{order}) 
which follow a Fibonacci pattern $LSLLSLSLLSLLS.....$. 
This results in three kinds of (effective) on-site potentials 
namely, $\epsilon_\alpha$, $\epsilon_\beta$ and $\epsilon_\gamma$, 
representing vertices flanked by $LL$, $LS$ or $SL$ bonds. 
The hopping integrals along the effective $L$ or $S$ bonds are designated by 
$t_L$ and $t_S$ respectively. These on-site potentials
\begin{figure}[ht]
\centering
\includegraphics[clip,width=5.5cm,angle=0]{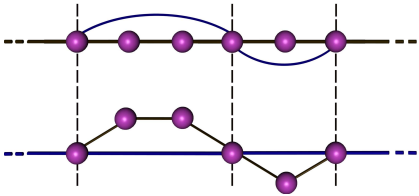}
\caption{(Color online) A one-dimensional chain of atoms with identical
atomic sites each of having on-site 
potentials $\epsilon$ and hopping integral $t$ with long range hopping 
$\lambda$ (blue colored line) across $n$-atoms and across 
$m$-atoms which follow a Fibonacci 
sequence, as discussed in text. In this particular case
we have taken $n=2$ and $m=1$ for instance.
The figure below indicates the equivalent \textit{decorated} version of
the buckled ordered chain with restricted long range hopping.} 
\label{order1}
\end{figure}
and the hopping 
integrals are given by,
\begin{eqnarray}
\epsilon_\alpha & = & \epsilon + 2 t \frac{U_{n-1}(x)}{U_n(x)} \nonumber \\
\epsilon_\beta & = & \epsilon + t \left [\frac{U_{n-1}(x)}{U_n(x)} + 
\frac{U_{m-1}(x)}{U_m(x)} \right ] \nonumber \\
\epsilon_\gamma & = & \epsilon_\beta \nonumber \\
t_L & = & \lambda + \frac{t}{U_n(x)} \nonumber \\
t_S & = & \lambda + \frac{t}{U_m(x)} 
\label{fibopara}
\end{eqnarray} 
Here, $x=(E-\epsilon)/2t$, and $U_n(x)$ is the $n$-th order 
Chebyshev polynomial of the second kind.
The resulting {\it effective} Fibonacci chain (Fig.~\ref{order}(b)) is then 
further renormalized using the standard 
decimation procedure, viz., by `folding' it backward using the 
deflation rule $LS \rightarrow L'$ and $L \rightarrow S'$. 
The recursion relations relating to the potentials and the hopping 
matrix elements at one length scale to the next, given by,
\begin{eqnarray}
\epsilon_\alpha' & = & \epsilon_\gamma + \frac{t_L^2 + t_S^2}{E - \epsilon_\beta} 
\nonumber \\
\epsilon_\beta' & = & \epsilon_\gamma + \frac{t_S^2}{E - \epsilon_\beta} \nonumber \\
\epsilon_\gamma' & = & \epsilon_\alpha + \frac{t_L^2}{E - \epsilon_\beta} \nonumber \\
t_L' & = & \frac{t_L t_S}{E - \epsilon_\beta} \nonumber \\
t_S' & = & t_L
\label{recur}
\end{eqnarray}
The above equations are exact.
\begin{figure}[ht]
\centering
\includegraphics[clip,width=8.5cm,angle=0]{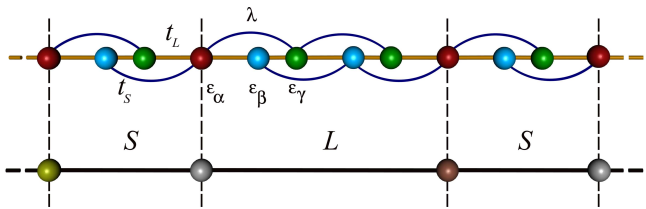}
\caption{(Color online) Schematic view of a part of an infinite Fibonacci
lattice comprising of long ($L$) and short ($S$) bonds having three different 
kinds of sites 
$\alpha$ (red sphere), $\beta$ (cyan sphere) and $\gamma$ (green sphere)
with corresponding on-site potentials 
$\epsilon_{\alpha}$, $\epsilon_{\beta}$ and $\epsilon_{\gamma}$ with
{\it restricted} short ranged $NNN$ interactions shown by blue curved lines.
The figure below indicates the \textit{renormalized} version of the
buckled Fibonacci chain.} 
\label{fibo}
\end{figure}
The spectrum is then easily obtained by calculating the density of states 
(DOS) by evaluating the Green's function 
$G_\alpha = (E + i\eta - \epsilon_\alpha^{*})^{-1}$ at any local site, say $\alpha$,
where, $\epsilon_\alpha^{*}$ is the fixed point value of the potential 
$\epsilon_\alpha$, 
which is achieved when the hopping integrals $t_L$ and $t_S$ flow to zero under 
iteration, 
and $\eta$ is a small imaginary part added to the energy. 
The DOS is given by, 
$\rho_\alpha = (-1/\pi) \lim_{\eta \rightarrow 0} Im G_\alpha (E + i\eta)$.

Before we end this sub-section, it is nice to appreciate that, the ``buckling'' 
simulates a long range hopping between sites that are not the 
nearest neighbors in the parent 1-D chain. The linear lattice depicted in 
Fig.~\ref{order} can be equivalently drawn as a quasiperiodic arrangement of 
say, squares and triangles (for $n=2$ and $m=1$) where the distant neighbors 
are connected by an overlap integral $\lambda$. A section of the original linear chain 
and its quasi-one dimensional equivalent is shown in Fig.~\ref{order1}. Thus the 
essence of ``buckling" is just to bring in the flavor of long range hopping in the 
system. Interestingly, similar long range hopping is considered recently in a Schr\"{o}dinger chain, and has been shown to 
lead to a new extremum at the center of the density of states accompanied by a van Hove singularity
~\cite{stock}.

\subsection{Quasiperiodic buckling of a quasiperiodic chain}
We consider a Fibonacci chain grown using two letters $L$ and $S$ following 
the growth rule $L \rightarrow LS$ and $S \rightarrow L$. A segment of an infinite 
Fibonacci chain is depicted in Fig.~\ref{fibo}. The $L$ and $S$ bonds are associated 
with the hopping integrals $t_L$ and $t_S$ respectively at the bare scale of length. 
The three different kinds of sites $\alpha$, $\beta$ and $\gamma$ lie between an 
$LL$, $LS$ and $SL$ pairs of bonds respectively. We consider buckling of the parent 
chain such that a tunnel hopping $\lambda$ is established across 
the combination of $LS$ or $SL$ 
pairs of bonds, but not across an $LL$ pair. Thus the present model can equivalently 
be looked at as a quasiperiodic chain with `restricted' second neighbor hopping.

The parent lattice with such restricted second neighbor hopping can be renormalized 
into a Fibonacci chain with nearest neighbor hoppings only by decimating a subset of 
vertices and now, using a different length scaling, viz., 
 $LSLSL \rightarrow L'$ and $LSL \rightarrow S'$.
\begin{figure}[ht]
\centering
\includegraphics[clip,width=8.5cm,angle=0]{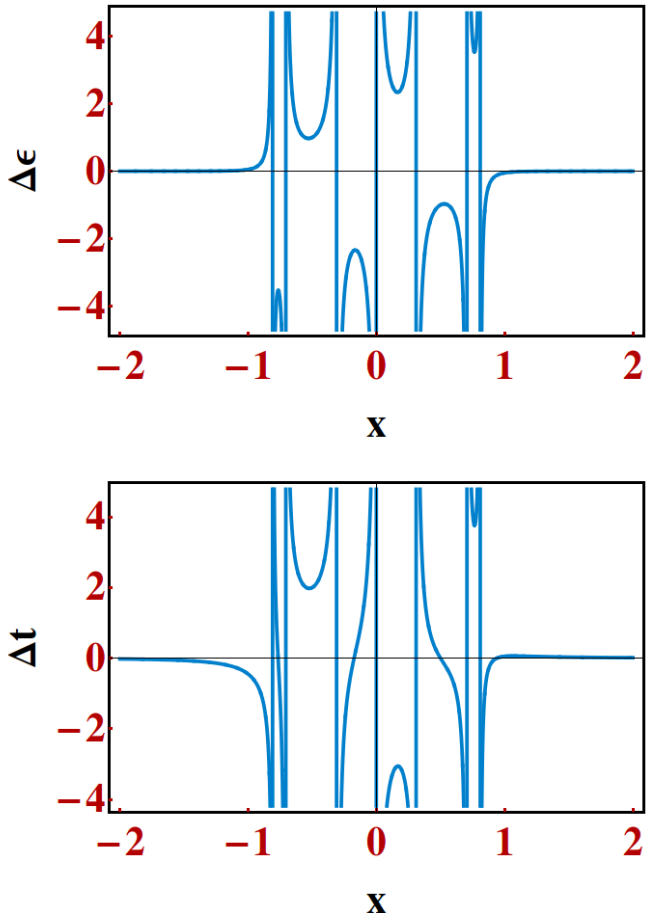}
\caption{(Color online) (a) $\Delta\epsilon = \epsilon_\alpha - \epsilon_\beta$ 
plotted against $x$. We have selected 
$\epsilon=0$ and $t=1$. $\epsilon_\alpha$, $\epsilon_\beta$, $t_L$ and $t_S$ are 
defined in Eq.~\eqref{fibopara} and (b) $\Delta t=t_L - t_S$ plotted against 
$x$.The dimensionless 
parameter $x=(E-\epsilon)/2t$ and the energy $E$ is measured 
in units of $t$.} 
\label{edif}
\end{figure}
To evaluate the DOS at any particular site one can use the same set of recursion 
relations given in Eq.~\eqref{recur}.
\section{Results and discussion}
\subsection{\textcolor{red}{The periodic chain case}}
Let us fix, for the sake of clarity, $\epsilon=0$ and $t=1$. 
Using the set of Eq.~\eqref{fibopara} we have worked out the difference
$\Delta\epsilon = \epsilon_\alpha - \epsilon_\beta$ and 
$\Delta t = t_L - t_S$ with the tunnel hopping being set in the 
range $-3 \le \lambda \le 3$. The 
results are displayed in Fig.~\ref{edif}, 
as a function 
of $x=(E-\epsilon)/2t$. It is seen that, for $-4 \le E \le -2$ and $2 \le E \le 4$, 
the differences $\Delta\epsilon$ and $\Delta t$ are both zero, implying that for such 
ranges of the energy the parameters represent a perfectly ordered chain.
Both the graphs show that the system `feels' disorder solely in the energy range $-2 \le E \le 2$.
Due to the specific pattern imposed in the distribution of the buckled segments, this 
part of the spectrum gets maximally perturbed by the aperiodic distribution of the 
tunnel hopping $\lambda$.
The existence of these localized states are marked by the divergence of $\Delta\epsilon$ 
at various values of $x$ within the range shown in the figure.
\begin{figure}[ht]
\centering
\includegraphics[clip,width=7cm,angle=0]{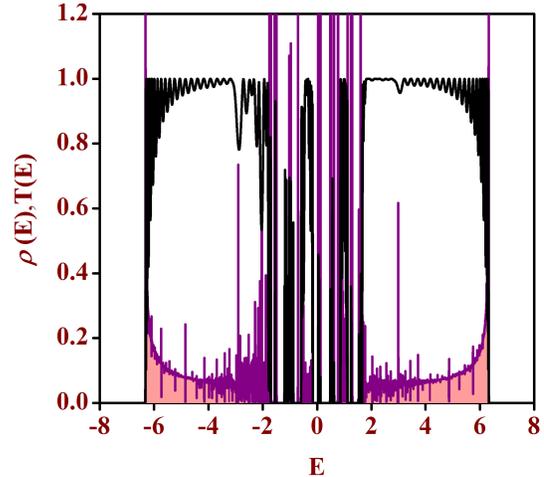}
\caption{(Color online) Density of states (violet curve) and two terminal 
electronic transport characteristics (black curve) of an ordered chain 
buckled sequentially at every $4$-th and $5$-th sites (that is, $n=3$ and $m=4$)
following a Fibonacci pattern. We have selected $\epsilon=0$, $t=1$ and 
$\lambda = 3$. Energy is measured in units of the hopping integral $t$.
The values of the lead parameters for the transport calculation
are $\epsilon_0 = 0$ and $t_0 = 3.5$.} 
\label{dospure}
\end{figure}
The number of subbands and 
such localized states depends strongly on the values of $n$, $m$ and of course, on the 
strength of the tunnel hopping amplitude $\lambda$. But, in every case we come across a scenario 
where a smooth crossover from an absolutely continuous part in the spectrum to a 
mixed character, dominated by subbands of narrower widths and point like spectrum, 
can be engineered.

The DOS, as presented in Fig.~\ref{dospure} exhibits an apparent continuum towards the 
outer parts of the spectrum with a typical divergence shown by an ordered lattice at the 
band edges. This can be understood in a perturbative way. For example, when $\lambda$ is 
large compared to $t$, we can, as a first approximation, ignore $t$ and the system 
resembles a linear chain with nearest neighbor hopping integral $\lambda$. The band extends from 
$\epsilon-2 \lambda$ to $\epsilon+2\lambda$ and the edges should exhibit a square root divergence 
as in an ordered chain of atomic sites. If we now `switch on' a non-zero value for $t$, the system 
starts feeling the presence of a quasiperiodic Fibonacci ordering of the potentials and the 
nearest neighbor hopping integrals. However, as long as $t$ is not comparable to $\lambda$ only the 
central part of the spectrum gets affected, while for $t \approx \lambda$, the entire spectrum gets 
affected, and gap opens everywhere.

As seen in Fig.~\ref{dospure} a gap opens up at 
the center of the spectrum, 
punctuated by tiny clusters of \textit{isolated} peaks. 
We have extensively investigated the flow pattern in such cases. Typically for any isolated peak 
in the density of states the number of RSRG iterations vary between $10$ to $15$. This means that 
the overlap of the Wannier orbitals extends typically over clusters of size $\tau^{10}$ to $\tau^{15}$.
Beyond this length the envelope decays down, but this indeed is a \textit{slow decay}. This scenario 
is to be contrasted with exponentially localized 
single particle states (not a member of the continuum regime),
for which the hopping integrals flow to zero very quickly (as the wave function 
of the electron is localized pretty sharply over a limited number of atomic sites)~\cite{south1,south2}.
We thus get confidence to comment that such isolated states are \textit{critical} in the 
usual sense of a Fibonacci lattice.

On the other hand, the shaded 
portions of the continuous subbands in the DOS 
are populated by extended eigenfunctions. For any energy eigenvalue picked up
at random from such continuous zones, the
nearest neighbor hopping integrals never flow to zero under the RSRG iterations. 
This proves that 
the envelopes of the wave functions have 
non-zero overlap at all scales of length. This means they are {\it extended}.

The black curve in Fig.~\ref{dospure} depicts the variation of two-terminal 
transmission coefficient as a function of the Fermi energy.  
To calculate the transmission coefficient, 
we have first placed a large but finite segment 
of the sample in between two perfectly ordered semi infinite `leads' where the atoms 
have a constant on-site potential $\epsilon_0$ and nearest neighbor hopping integral 
$t_0$. The ordered chain with the quasiperiodic buckling is transformed in to an 
`effective' Fibonacci chain. The transmission coefficient is then worked out using 
the well known formula~\cite{stone}, viz., 
\begin{widetext}
\begin{equation}
T(E) = \frac{4 \sin^2 ka}{[(M_{12}-M_{21}) + (M_{11}-M_{22}) \cos ka)]^2 + 
[(M_{11}+M_{22})^2 \sin^2 ka]}
\label{tran}
\end{equation}
\end{widetext}
where, $\cos ka = (E-\epsilon_0)/2t_0$, $M = M_R (\prod M_{i}) M_L$, 
with $M_i = M_\alpha$, 
$M_\beta$ or $M_\gamma$ following the Fibonacci sequence generated~\cite{kohmoto1}, and 
$M_{R(L)}$ being the transfer matrix corresponding to the atoms sitting at the 
right and the left extremities respectively. 
The explicit form of the matrices are,
\begin{widetext}
\begin{equation}
M_\alpha = \left( \arraycolsep=5pt \def\arraystretch{1.5} \begin{array}{cccc}
\frac{(E-\epsilon_\alpha)}{t_L} & -1 \\ 
1 & 0 
\end{array}
\right), 
M_\beta = \left( \arraycolsep=5pt \def\arraystretch{1.5} \begin{array}{cccc}
\frac{(E-\epsilon_\beta)}{t_S} & -\frac{t_L}{t_S} \\ 
1 & 0 
\end{array}
\right),
M_\gamma = \left( \arraycolsep=5pt \def\arraystretch{1.5} \begin{array}{cccc}
\frac{(E-\epsilon_\gamma)}{t_L} & -\frac{t_S}{t_L} \\ 
1 & 0 
\end{array}
\right)
\end{equation}
\end{widetext}
The `left' and the `right' matrices are also given by,
\begin{equation}
M_L = \left( \arraycolsep=5pt \def\arraystretch{1.1} \begin{array}{cccc}
\frac{(E-\epsilon_L)}{t_S} & -\frac{t_0}{t_S} \\ 
1 & 0 
\end{array}
\right),
M_R = \left( \arraycolsep=5pt \def\arraystretch{1.1} \begin{array}{cccc}
\frac{(E-\epsilon_R)}{t_0} & -\frac{t_L}{t_0} \\ 
1 & 0 
\end{array}
\right)
\end{equation}
Here $\epsilon_L$ and $\epsilon_R$ are the on-site potentials of
the left and right atoms of the finite size atomic chain respectively.
\subsection{The Fibonacci chain}
We refer to Fig.~\ref{fibo}. The infinite Fibonacci chain is buckled 
so that a tunnel hopping $\lambda$ connects sites separated by an $LS$, or 
an $SL$ pair, but excludes an $LL$ pair. This geometry equivalently 
represents a Fibonacci chain with `restricted second neighbor hopping'. It is 
interesting to observe that such a model can be mapped onto a nearest neighbor 
hopping model by a deflation rule, $LSLSL \rightarrow L'$ and $LSL \rightarrow S'$. 
This decimates a subset of sites in the original chain to yield a renormalized 
version of it. The effective on-site potentials and hopping integrals for this 
renormalized Fibonacci lattice can be worked out in a tedious, but straightforward 
manner. For clarity we present below the initial values of the 
parameters for the effective 
Fibonacci chain with nearest neighbor interaction. 
To simplify the cumbersome expressions we select 
$\epsilon_\alpha=\epsilon_\beta=\epsilon_\gamma=0$ in the original chain. This doesn't 
affect the physics as the choice of the on-site potentials only sets the centre 
of the density of states spectrum. For the `new' Fibonacci chain the parameters 
read, 
\begin{equation}
\begin{aligned}
& \epsilon_{\alpha,new} = \frac{{\cal F}_1 E^3 + {\cal F}_2 E^2 + 
{\cal F}_3 E + {\cal F}_4}
{E^4 + {\cal G}_1 E^2 + {\cal G}_2 E + {\cal G}_3} \\
& \epsilon_{\beta,new} = \frac{{\cal F}_5 E^5 + {\cal F}_6 E^4 + {\cal F}_7 E^3 + 
{\cal F}_8 E^2 + {\cal F}_9 E}{-E^6 + {\cal G}_4 E^4 - {\cal G}_2 E^3 + {\cal G}_5 E^2 + 
{\cal G}_6 E + {\cal G}_7 } \\
& \epsilon_{\gamma,new} = \epsilon_{\beta,new} \\
& t_{L,new} = \frac{{\cal F}_{10} E^2 + {\cal F}_{11} E + 
{\cal F}_{12} + {\cal F}_{13}}
{E^4 + {\cal G}_8 E^2 + {\cal G}_2 E +{\cal G}_9} \\
& t_{S,new} = \frac{2 \lambda t_L E + t_S (\lambda^2 + t_L^2)} 
{E^2 - t_S^2}.
\end{aligned}
\label{newfibo}
\end{equation}
where, ${\cal F}_1 = 2 (t_L^2 + \lambda^2)$, ${\cal F}_2 = 4\lambda t_L t_S$, 
${\cal F}_3 = 2 [ t_L^2 ( \lambda^2 - t_L^2 - t_S^2) -\lambda^2 (\lambda^2 + t_S^2)]$, 
${\cal F}_4 = 4 \lambda t_L t_S (\lambda^2 -t_L^2 - t_S^2)$, 
${\cal F}_5 = -2 (\lambda^2 + t_L^2)$, ${\cal F}_6 = -4 t_L t_S \lambda$, 
${\cal F}_7 = 2 t_L^4 + 4 t_S^2 (\lambda^2 + t_L^2) + \lambda^2 (2 t_L^2 + 3 \lambda^2)$, 
${\cal F}_8 = 8 \lambda t_L t_S (t_L^2 + t_S^2) + 6 t_L t_S \lambda^3$, 
${\cal F}_9 = \lambda^4 (t_S^2 - t_L^2) - \lambda^6 - t_L^2 t_S^2 (t_L^2 + 11 \lambda^2) - 2 
t_S^4 (t_L^2 + \lambda^2)$, 
${\cal F}_{10} = 3 t_L \lambda^2$,
${\cal F}_{11} = 2 \lambda t_S (2 t_L^2 + \lambda^2)$,
${\cal F}_{12} = 2 \lambda^2 t_L (t_S^2 - \lambda^2)$,
${\cal F}_{13} = t_L^3 (t_S^2 + \lambda^2)$,
${\cal G}_1 = -(t_L^2 + 2 t_S^2 + 2 \lambda^2)$, ${\cal G}_2 = -4 \lambda t_L t_S$, 
${\cal G}_3 = \lambda^4 - t_S^2 (2\lambda^2 - t_S^2)$, 
${\cal G}_4 = t_L^2 + 3 t_S^2 + 2 \lambda^2$, ${\cal G}_5 = -[\lambda^4 + 
t_S^2 (t_L^2 + 3 t_S^2 + 4 \lambda^2)]$, ${\cal G}_6 = -4 \lambda t_L t_S^3$,  
${\cal G}_7 = t_S^6 - \lambda^2 t_S^2 (\lambda^2 + 2 t_S^2)$,
${\cal G}_8 = - (2 t_S^2 + 2 \lambda^2 + t_L^2)$, and
${\cal G}_9 = (t_S^2 - \lambda^2)^2$.

\begin{figure}[ht]
\centering
\includegraphics[clip,width=8.5cm,angle=0]{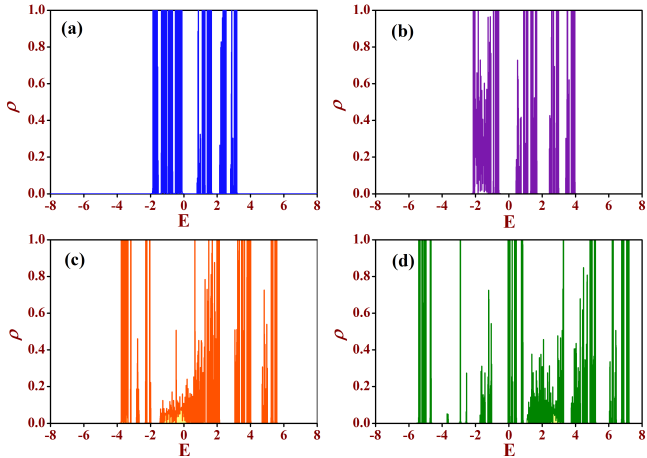}
\caption{(Color online) Density of states of an infinite Fibonacci 
chain as a function of the energy for different combination of
buckling constant $\lambda$.
The buckling has been applied at the bare length scale.
We have set $\epsilon_i=0$, $i=\alpha$, 
$\beta$ and $\gamma$. $t_L=1$, $t_S=1.5$.
Energy is measured in 
units of $t_L$. The values of the buckling parameters in
the following four cases respectively are
(a) $\lambda=0.5$
(b) $\lambda=1.0$
(c) $\lambda=2.0$
(d) $\lambda=3.0$. } 
\label{fibo1}
\end{figure}

One can now use Eq.~\eqref{recur} with the initial values of the on site 
potentials and the nearest neighbor hopping integrals as listed above to work out the 
density of states, as discussed before.
The density of states at any site of this renormalized, effectively nearest neighbor 
Fibonacci chain can be obtained using the recursion relations Eq.~\eqref{recur}. The 
local density of states at an $\alpha$-site is presented in Fig.~\ref{fibo1}(a), when 
we introduce the tunnel hopping $\lambda$ at the bare scale of length. This 
implies that the Fibonacci chain is wrinkled every $LS$ or $SL$ pair. The 
effect on the spectrum is most severe in this case. However, surprisingly, 
patches of continuous distribution of eigenvalues turn out to be an 
interesting feature in this case. 

In Fig.~\ref{fibo1} we show sequentially the 
effect of variation of the long range hopping on the density of states. $\lambda$ 
is chosen to have values $0.5$, $1.0$, $2.0$ and $3.0$ in Fig.~\ref{fibo1} (a), (b), 
(c) and (d) respectively. While for $\lambda=0.5$ the density of states does not show any 
significant continuum, for $\lambda=1.0$, a trace of a continumm appears to the left of 
$E=0$. The continuum appears much more prominently as $\lambda$ increases to $2.0$ and 
$3.0$ respectively. For each such case, we have tested the flow of the hopping integrals
under successive renormalization. The hopping integrals, for any energy 
eigenvalue picked at random from such dense patches of energy, keep on oscillating without 
converging to zero, indicating that the states are extended in character.
We have thoroughly examined such regimes with finer and finer scanning of the 
intervals of energy. 

\begin{figure}[ht]
\centering
\includegraphics[clip,width=8.5cm,angle=0]{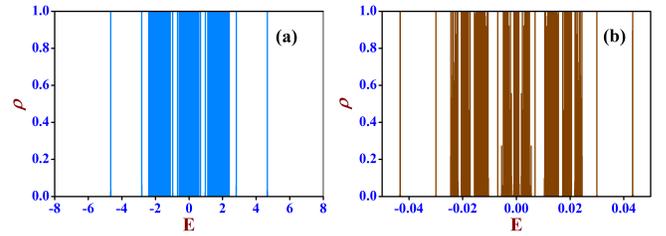}
\caption{(Color online) Density of states of an infinite Fibonacci 
chain. 
(a) Buckling is applied to a $20$ times renormalized 
Fibonacci chain. (b) Enlarged central part of (a) to bring out the 
inherent self similarity displayed by the distribution of the eigenvalues.
We have set $\epsilon_i=0$, $i=\alpha$, 
$\beta$ and $\gamma$. $t_L=1$, $t_S=1.5$ and $\lambda=2.5$. Energy is measured in 
units of $t_L$.} 
\label{fibo2}
\end{figure}

In Fig.~\ref{fibo2} we impart buckling on a $20$ times 
renormalized Fibonacci chain to begin with. Naturally, such a 
long chain retains the essential features of an infinite lattice.
The figure shows the three subband structure in the 
DOS pattern, a trademark of the transfer model of a Fibonacci chain~\cite{kohmoto2}.
However, the bound states caused by the buckling are still observed within the 
spectral gaps of the three subband area, as well as outside the global band edges.

The fine structure of the spectrum is brought out as shown in Fig.~\ref{fibo2}(b) 
around the central part. The typical self similar three subband spectrum opens up, with 
further splitting being visible. The additional feature is the appearance of 
sharply localized bound states inside the multiple gaps in the spectrum, indicating 
a self similar distribution of the bound states as well. 

\begin{figure}[ht]
\centering
\includegraphics[clip,width=8.5cm,angle=0]{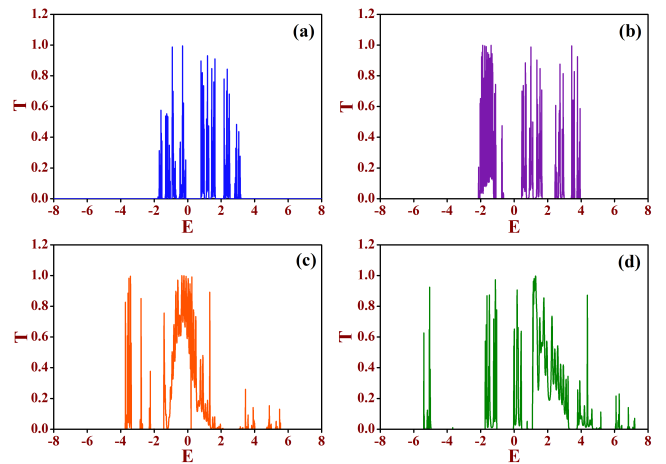}
\caption{(Color online) Two-terminal transmission characteristics
 of a Fibonacci 
chain with $144$ bonds as a function of the energy for different combination of
buckling constant $\lambda$.
The buckling has been applied at the bare length scale.
We have set $\epsilon_i=0$, $i=\alpha$, 
$\beta$ and $\gamma$. $t_L=1$, $t_S=1.5$.
Energy is measured in 
units of $t_L$. The values of the buckling parameters in
the following four cases respectively are
(a) $\lambda=0.5$
(b) $\lambda=1.0$
(c) $\lambda=2.0$
(d) $\lambda=3.0$.}
\label{transfibo}
\end{figure}

The two terminal transport across a $144$-bond long Fibonacci chain, 
buckled at the bare length scale with the strength of the 
tunnel hopping $\lambda$ following the sequence of values as 
in Fig.\ref{fibo1}, is shown in Fig.~\ref{transfibo}. The parameters 
are the same as those in Fig.~\ref{fibo1}. The absolutely continuous character 
present in the density of states spectrum is corroborated by a continuous zone of
high transmission coefficient which certify for the extended character of the spectrum.

\begin{figure}[ht]
\centering
\includegraphics[clip,width=8.5cm,angle=0]{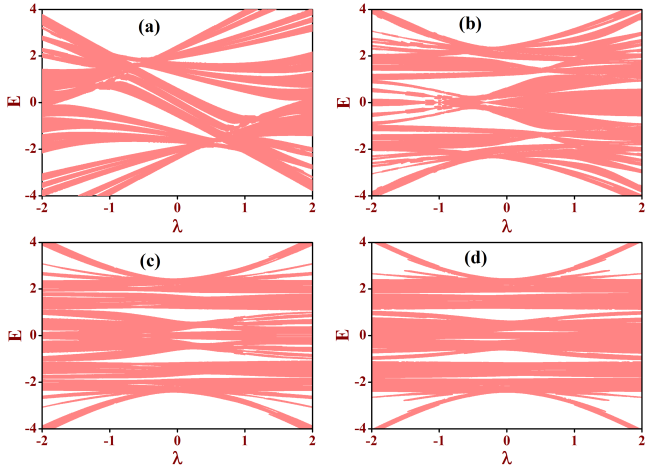}
\caption{(Color online) $E$ versus $\lambda$ profile of a Fibonacci lattice at 
different stage of initial renormalization (a) $n=0$, (b) $n=2$, (c) $n=4$ and
(d) $n=6$.
The parameters are
$\epsilon_{\alpha}=\epsilon_{\beta}=\epsilon_{\gamma}=0$ and $t_{L}=1$ and $t_{S}=1.5$.} 
\label{fibospec}
\end{figure}

To end, we find it instructive to illustrate the distribution of the energy eigenvalues 
of a buckled Fibonacci chain as a function of the tunnel hopping amplitude $\lambda$.
Fig.~\ref{fibospec} displays the effect of buckling. The four panels correspond to 
the different scales of length at which the buckling has been introduced. Panel (a) 
shows the effect when the chain is bent at its bare scale of length. 

In general, the energy spectrum of a Fibonacci quasicrystal, in an 
off-diagonal model, consists of three fragmented subbands, each splitting 
further into three self similar subbands on finer scan over the 
energy interval~\cite{kohmoto1}. In the present case, with buckling 
at the bare scale of length we can see that 
this three subband pattern is destroyed. 
The effect of the buckling is maximum here. 
The bound states 
arising out of the sequential bending are not visible prominently, but the global 
three subband spectral splitting of a Fibonacci chain with nearest neighbor interaction 
is lost. With the buckling introduced over longer segments, achieved by renormalizing 
the original chain, say, two, four or six times first, the three subband pattern 
seems to be getting restored gradually, and is quite apparent in the last of 
the panels, that is, (d). This is understandable, as the introduction of the buckling 
through a longer range tunnel hopping on an $n$-step renormalized lattice implies that 
the nearest neighbor quasiperiodic chain's spectrum is preserved as we have larger 
and larger segments of `clean' Fibonacci lattice trapped in the {\it bent zone} 
for increasing values of $n$. The appearance of the bound states are also apparent 
as $\lambda$ increases. 
\section{conclusion}
We have addressed the problem of the changes brought about in the electronic spectra 
of a perfectly periodic 
chain of atoms and in a quasiperiodic Fibonacci chain when finite segments of them 
are buckled to bring two distant neighbors into close proximity. This `proximity' 
is modeled by a tunnel hopping connecting these distant neighbors. We have used a 
real space renormalization group decimation scheme to map the chains with restricted 
long range hoppings on to chains with renormalized on-site potentials and nearest neighbor 
hopping integrals, both being functions of the energy of the electron. It is seen that 
the buckling induced spectral characters of the pure and the quasiperiodic chain 
are quite contrasting. With large values of the tunnel hopping integral, a measure 
of the amount of buckling, the `ordered' chain can retain its absolutely continuous 
part towards the end of the spectrum, populated by extended 
eigenstates only, while the central part turns `dirty' and offers a 
mixed character of extended and localized states. This can be utilized to look for 
a possible `switch on', `switch off' effect in its electronic conduction.

The Fibonacci chain on the other hand, shows the startling appearance of one or 
more {\it absolutely continuous} subbands populated 
by unscattered (extended) eigenstates flanked by the usual {\it critical} ones.
The appearance of sharply localized, bound states mark both the case, particularly 
when the buckling is large, that is the tunnel hopping integral $\lambda$ assumes 
a large value compared to the intrinsic hopping parameters of the systems. Such states 
appear to get 
distributed in a self similar manner, a fact that becomes more apparent when 
the buckling is introduced over larger and lager segments of the original 
Fibonacci chain.

It is to be appreciated that the long range hopping brings in a flavour of the formation 
of \textit{rings}, providing additional path for the electron to jump to a distant site.
The `buckling' essentially gives rise to the formation of polygonal structures (see 
Fig.~\ref{order1}). For example, with $n=2$ and $m=1$ one encounters a sequence 
of trapezia and triangles, the latter creating frustration and the anti-bonding states 
gets pushed away in energy. This has immense effect on the localization properties 
~\cite{naumis2}. Such buckling has been implemented experimentally in graphene nanoribbons 
very recently~\cite{lit} for both in-plane and out-of-plane bendings, and 
the changes in the density of states have been probed. 
The spectrum is not unlikely to display similar or even richer properties if one studies 
a two dimensional strip, buckled and stressed. This will be addressed theoretically in a future communication.
\begin{acknowledgments}
\textcolor{red}{A.M. acknowledges DST, India for the financial support provided through an INSPIRE Fellowship [IF$160437$].} A.N. is thankful for the financial support provided through a research fellowship
[Award letter no. F.$17$-$81 / 2008$ (SA-I)] from UGC, India.
A.C. acknowledges partial financial support from a DST-PURSE grant through the 
University of Kalyani.
\end{acknowledgments}

\end{document}